# A Comparative Study of Digital Memristor-Based Processing-In-Memory from a Device and Reliability Perspective

*Thomas Neuner, Henriette Padberg, Lior Kornblum, Eilam Yalon, Pedram Khalili Amiri, and Shahar Kvatinsky\**

As data-intensive applications increasingly strain conventional computing systems, processing-in-memory (PIM) has emerged as a promising paradigm to alleviate the memory wall by minimizing data transfer between memory and processing units. This review presents the recent advances in both stateful and non-stateful logic techniques for PIM, focusing on emerging nonvolatile memory technologies such as resistive random-access memory (RRAM), phase-change memory (PCM), and magnetoresistive random-access memory (MRAM). Both experimentally demonstrated and simulated logic designs are critically examined, highlighting key challenges in reliability and the role of device-level optimization in enabling scalable and commercial viable PIM systems. The review begins with an overview of relevant logic families, memristive device types, and associated reliability metrics. Each logic family is then explored in terms of how it capitalizes on distinct device properties to implement logic techniques. A comparative table of representative device stacks and performance parameters illustrates trade-offs and quality indicators. Through this comprehensive analysis, the development of optimized, robust memristive devices for next-generation PIM applications is supported.

## 1. Introduction

The end of Dennard scaling[1] has heightened the challenge of achieving energy efficiency in computing systems, resulting in most modern computers approaching their performance limits.[2] With the slowing of Moore's law and the growing impact of the memory wall[3]—the widening gap between CPU speed and memory response time—identifying the next technological breakthrough has become critical. Today's computers still rely on the von Neumann architecture, which separates processing and memory units and have benefited from this design for over 70 years. However, the inability to further reduce data transfer between these units poses a severe bottleneck to both speed and energy efficiency.[4] Consequently, non-von Neumann architectures are attracting significant research interest.

One promising approach is *neuromorphic computing*,[5] which takes inspiration from the structure and functionality of the brain by employing neurons and synapses to execute neural network models. Beyond neuromorphic systems, another important paradigm challenging the von Neumann hegemony is *Processing-in-Memory* (PIM).[6] In PIM, logic operations are conducted directly within the memory unit, thereby minimizing data transfer and alleviating the von Neumann bottleneck.[7]

Within this broad landscape, this paper focuses on digital PIM based on logic gates constructed from memristive nonvolatile memory cells. A key requirement for such PIM approaches is a switching device that enables both data storage and computation in the same physical entity. Memristive devices, which can retain a resistive or memory state, fulfill these criteria by offering fast switching and energy-efficient computing.[8] Depending on the materials and their switching mechanisms, a range of logic techniques and technology combinations become possible.

This review presents current digital logic techniques on memristive technologies and provides an overview of state-of-the-art research. We first classify the logic techniques into *stateful* and *non-stateful* logic families. Then, we introduce the most prominent memristive technologies—resistive RAM (RRAM), phase change memory (PCM), and magnetoresistive RAM (MRAM)—along with their switching mechanisms and key reliability considerations. We subsequently discuss the fundamental principles

T. Neuner, P. K. Amiri
Department of Electrical and Computer Engineering
Northwestern University
Evanston, IL 60208, USA
H. Padberg, L. Kornblum, E. Yalon, S. Kvatinsky
Andrew & Erna Viterbi Faculty of Electrical and Computer Engineering
Technion – Israel Institute of Technology
Haifa 3200003, Israel
E-mail: shahar@ee.technion.ac.il

**DOI: 10.1002/aelm.202500348**

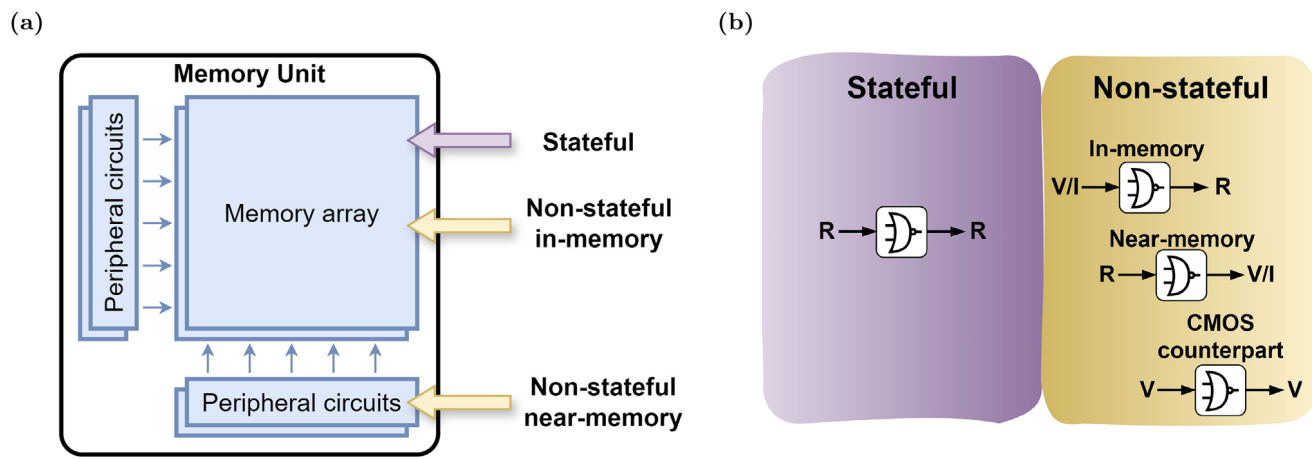

**Figure 1.** a) Illustration of a memory unit. Both stateful and non-stateful in-memory computing are executed within the memory array, while for non-stateful near-memory additional peripheral circuits, such as sense amplifiers (SA) are required.[141] b) Simplified categorization of stateful and non-stateful logic for memristors distinguishable by their typical input and output representations (R: resistance; V: voltage; I: current). Note that only the most popular input/output representations are presented here, as variants from this representation exist in some logic techniques. The symbol in the box represents an arbitrary logic gate (displayed here as a NOR gate).

of each logic technique and examine how different memristive switching mechanisms align with these approaches to pinpoint the gaps and opportunities for future works. We conclude by summarizing the devices and materials used for experimental demonstrations of these logic techniques. This paper serves as a call for action, encouraging further research on various memristive devices and advocating for more extensive experimental validation of PIM-based logic techniques.

## 2. Classification of Logic Families

Logic families provide a structured way of categorizing circuit components based on their capability to execute fundamental logic functions such as AND, OR, and NOR. Memristive logic circuits, in particular, can be characterized by several factors: the number of elements used in a logic operation, their interconnection pattern, the physical representation of the logic states (such as voltage or resistance), the coverage of basic logic operations, and their compatibility with crossbar arrays and other technologies (e.g., CMOS). Generally, memristive logic families are classified into two main categories: *stateful* logic and *non-stateful* logic.

**Figure 1a** illustrates how these logic families differ from an architectural standpoint, describing where the logic operation takes place and which parts of the computer architecture are employed. From the system-level perspective, there is no distinction between stateful and non-stateful in-memory computing. However, the difference becomes apparent when reading logical values and performing multiple logic operations.

The categorization of stateful and non-stateful logic techniques with respect to their input and output representation is illustrated in Figure 1b. A logic family is termed *stateful* if both its inputs and outputs share the same physical representation as the stored data with no data conversion necessary; for memristors, this representation is resistance, for static RAM (SRAM) it is voltage, for dynamic RAM (DRAM) it is charge. Conversely, non-stateful logic usually employs different representations for inputs and outputs. To perform a logic operation in non-stateful logic, data must be read or written - a process that involves converting between two different physical representations such as converting two input voltages into a resistance.

Within non-stateful logic, we can further categorize three subclasses based on where the computation takes place and the level of compatibility with crossbar/memory structure: *non-stateful in-memory*, *non-stateful near-memory*, and *non-stateful CMOS counterpart*. In both in-memory and near-memory approaches, at least one logic operand is stored as a resistance state, while the other is represented as a current or voltage, necessitating data conversion.

In *non-stateful in-memory* logic, the computation is confined strictly to the memory array. In contrast, *non-stateful near-memory* logic extends computation beyond the array by actively involving the peripheral circuits—such as sense amplifiers, drivers, or custom blocks—in addition to memory cells.

In the memristive *non-stateful CMOS counterpart*, memristive devices are integrated with CMOS transistors, usually in dedicated logic circuits outside the memory array. These configurations operate similarly to conventional transistor-based logic gates, with both inputs and outputs represented as voltages, enabling compatibility with standard CMOS logic. Architecture and circuit level aspects of these logic families were thoroughly reviewed by others,[9] covering the classification of PIM by evaluating the needed number of devices and the correlation between input and output cells.

The focus of this current review is the suitability of different device technologies for each logic technique evaluating material characteristics and various design considerations.

## 3. Memristive Devices for Logic

A variety of memristive technologies are available, each relying on a distinct physical mechanism.[10] This section reviews three memristive device types used for logic in memory: resistive random-access memory (RRAM),[11–13] phase change memory (PCM),[14,15] and magnetic RAM (MRAM).[16] The basic structure of each device is illustrated in **Figure 2**. Within RRAM, we discuss types of devices based on the valence change mechanism (VCM)[17,18] and electrochemical metallization (ECM),[19] whereas MRAM devices typically employ spin-transfer torque (STT)[20,21] or spin-orbit torque (SOT)[22] for switching. The following subsections briefly describe the fundamental switching principles of these technologies.

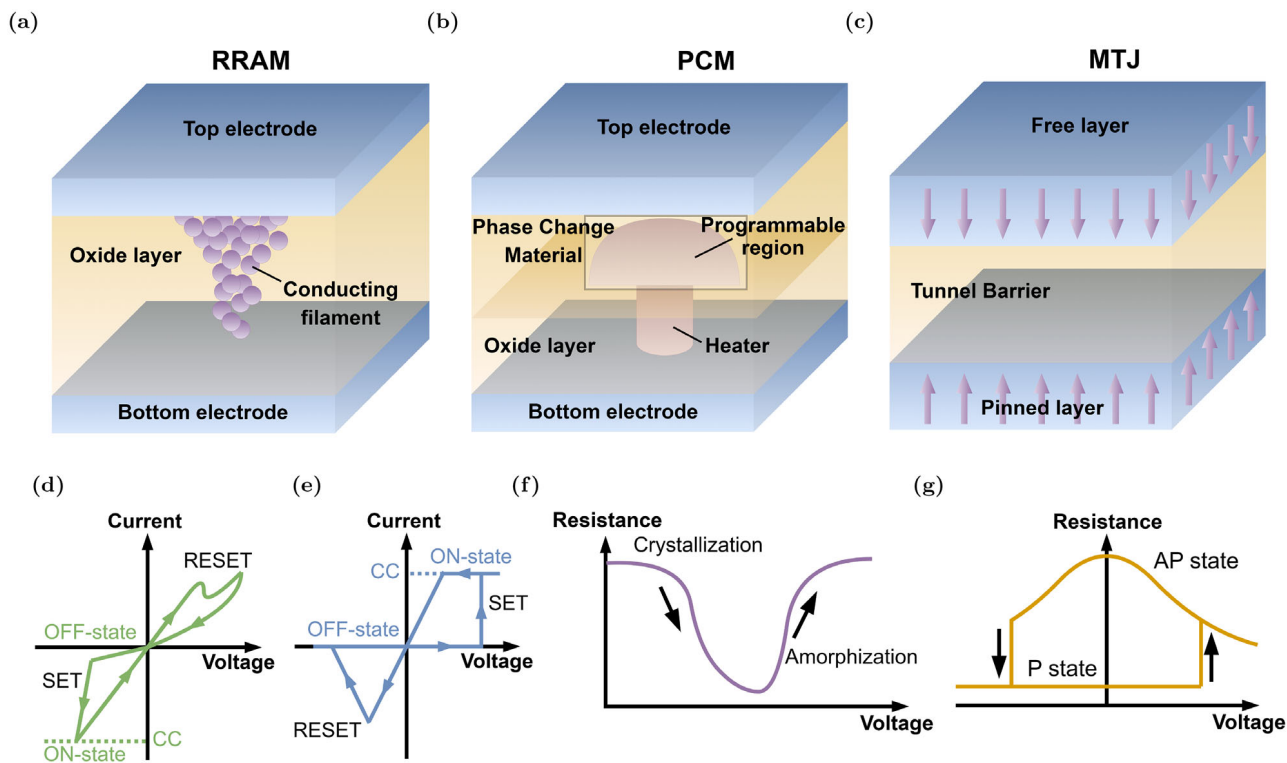

**Figure 2.** Illustration of the emerging memory technologies discussed in this work, along with their representative current–voltage (I–V) and resistance–voltage (R–V) characteristics. a) RRAM, featuring a typical metal–insulator–metal (MIM) structure and operating based on filamentary switching. b) PCM, utilizing a mushroom-type cell geometry, stores data by switching between crystalline and amorphous phases. c) MTJ in the anti-parallel configuration, consisting of a pinned and a free magnetic layer, allowing magnetization reversal in the free layer. d) I–V characteristic of a VCM cell (CC: Current Compliance). e) I–V characteristic of an ECM cell (CC: Current Compliance). f) R–V characteristic of a PCM cell. g) R–V characteristic of an STT-based MTJ cell, showing the resistance change between anti-parallel and parallel states.

## 3.1. Resistive RAM (RRAM)

Here we briefly review the switching mechanisms of the key types of RRAM devices. For a concise recent review of the mechanisms and materials we recommend ref. [18].

### 3.1.1. Valence Change Mechanism (VCM)

A VCM[17] memristive cell operates under *bipolar switching*, meaning that the polarity of the applied voltage (positive or negative) determines the cell's transition. As shown in Figure 2d, two metal electrodes (top and bottom) sandwich an insulator layer (or a layer stack) to form a switchable cell. The switching mechanism relies on the movement of oxygen vacancies under an applied voltage as depicted in **Figure 3a**. During a SET operation, the device transitions from an OFF state (high resistive state, HRS) to an ON state (low resistive state, LRS) by driving oxygen vacancies into a region that lowers the overall resistance. Most VCM cells require a current compliance to limit the current during the SET operation, thereby preventing hard breakdown, ensuring device functionality. Reversing the polarity initiates the RESET process, which returns the device to the high resistive state.

### 3.1.2. Electrochemical Metallization Memory (ECM)

Another RRAM technology is ECM,[19] also known as the conductive bridge RAM (CBRAM) or programmable metallization cells (PMCs). Similar to VCM, ECM exhibits bipolar switching and is driven by an applied electric field that oxidizes metal ions from an electrochemically active electrode which then migrate and form a conducting filament at the inert electrode as described in Figure 3b.[23] Similar to VCM devices a forming step is required to form the filament. As illustrated in Figure 3e, reversing the voltage polarity dissolves or diminishes this filament, thereby increasing the resistance and returning the cell to an HRS.

## 3.2. Phase Change Memory (PCM)

PCM[14,15] relies on the unique behavior of phase change materials (commonly chalcogenides) that transition between amorphous and crystalline phases (see Figure 3c,d). The transition is triggered by Joule heating as current flows through the device, followed by controlled cooling. Slow cooling below the melting temperature $T_{melt}$, but above the crystallization temperature $T_{crystal}$ yields to an ordered, crystalline phase with low resistance. In contrast, rapid cooling—shorter than the crystallization time (melt-quench)—from a molten state above $T_{melt}$ forms an amorphous phase with high resistance. A typical PCM temperature vs time switching characteristic for a mushroom-type cell is shown in Figure 3e. This technology is classified as *unipolar switching* because the relevant pulses differ primarily in their amplitude, rather than in polarity. This is also demonstrated in Figure 2f where small applied voltages cause a reduction in resistance (crystallization) and high voltages lead to an increase in resistance (amorphization).

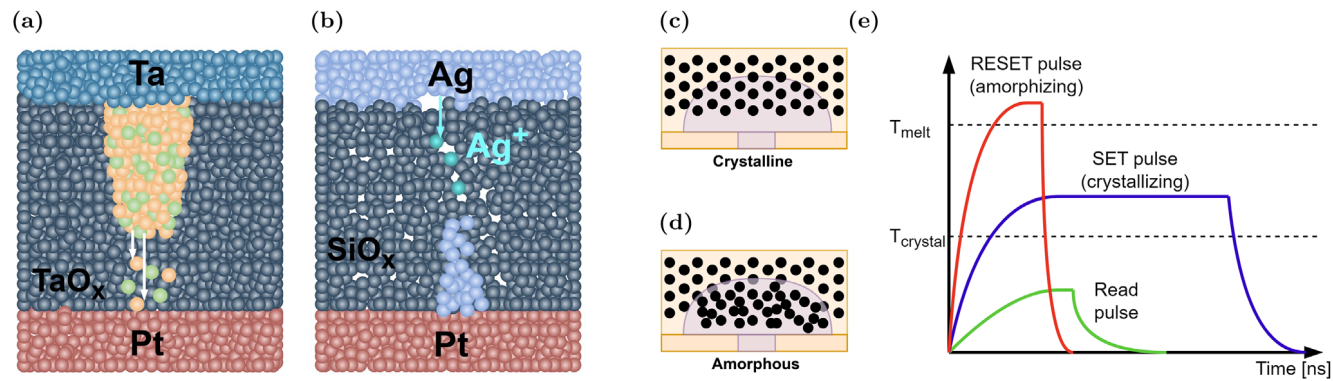

**Figure 3.** Schematic switching mechanism of RRAM and PCM devices. a) VCM mechanism showing the SET process. The green spheres represent oxygen vacancies ($V_O^-$), while the orange spheres represent metal ions in a lower valence state. Only the oxygen vacancies are mobile, while the metal ions can only change their valence state. In the RESET process, the filament is disrupted. b) ECM mechanism showing the SET process with ions starting to move from the active electrode (here Ag) to the electrochemically inert counter electrode (here Pt) and building a conductive filament. In the RESET process, the filament is disrupted. c) and d) Schematic of a mushroom-type PCM programming region showing the two different states, either crystalline (c) or amorphous (d). e) Applied electrical pulses to change the temperature in the phase-change layer. A fast pulse induces high temperature to reset the cell into the high resistive amorphous state. The SET pulse is a longer pulse that generates a medium temperature to crystallize the cell into a low-resistive state. The read pulse is performed at a lower voltage inducing negligible heating, so the state is not changed.

### 3.3. Magnetic Tunnel Junction (MTJ)

The magnetic tunnel junction (MTJ) is the core device used in MRAM technologies. An MTJ consists of two ferromagnetic layers separated by a thin insulating tunneling barrier. One ferromagnetic layer is fixed (pinned layer, PL), while the other (free layer, FL) is free to switch its magnetic orientation as in Figure 2c. As illustrated in **Figure 4**a,b, the MTJ exhibits different resistance values depending on whether the magnetic orientations of the two layers are parallel (P, low resistance) or anti-parallel (AP, high resistance). These two states arise due to the tunneling magnetoresistance (TMR) effect in sufficiently thin barrier layers.[24] Two key mechanisms, STT and SOT, used in logic operations, switch the FL orientation while the voltage-controlled

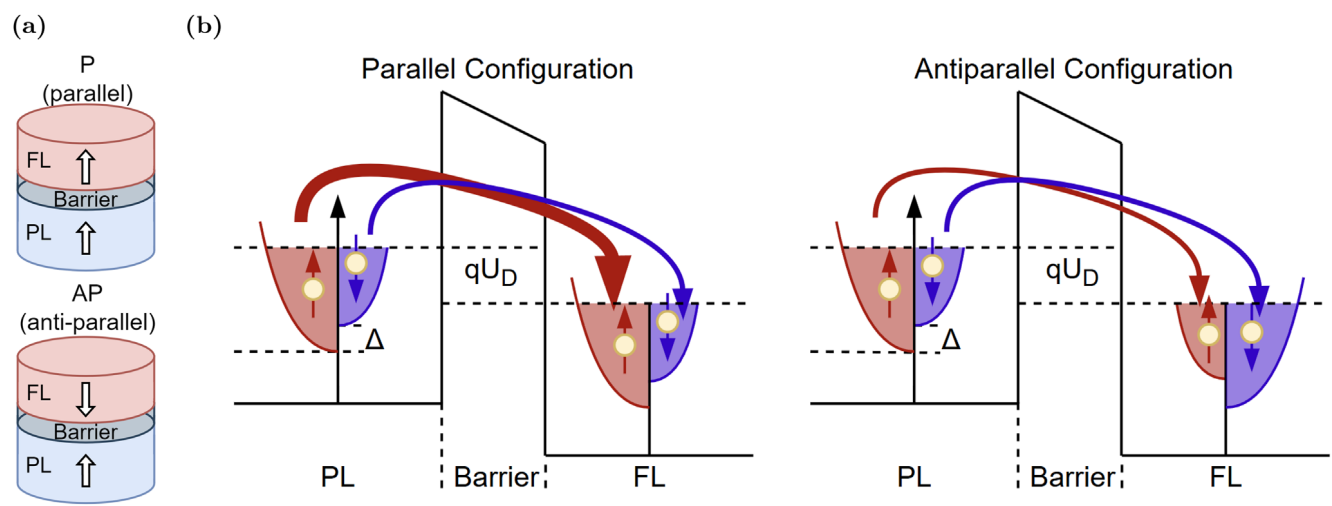

**Figure 4.** Perpendicular MTJ structure and TMR (Tunneling Magnetoresistance) effect used for the readout. a) The two states (parallel, P, and anti-parallel, AP) of a perpendicular MTJ (PL: Pinned Layer; FL: Free Layer). b) Energy band diagrams for the parallel (left) and anti-parallel (right) configuration of two ferromagnetic layers sandwiching an insulating layer. An applied voltage leads to a shift of the spin-up (blue) and spin-down (red) states and results in different partial tunneling through the insulator for P and AP state. Tunneling can only take place between the same spin-state. The spin splitting $\Delta$, and barrier height of the tunneling barrier ($qU_D$) are relatively indicated.

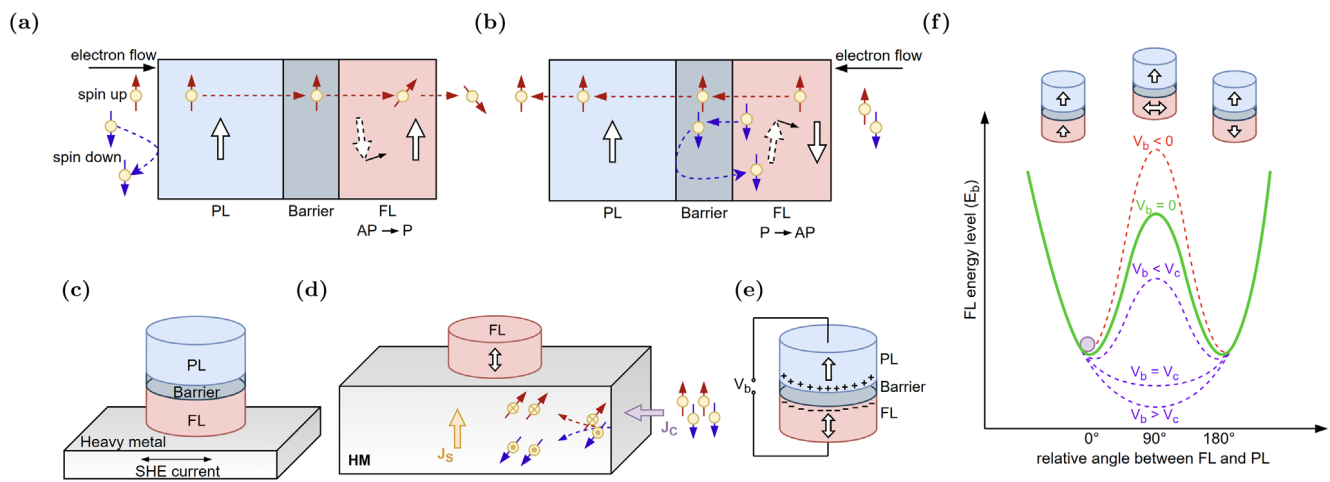

**Figure 5.** a,b) STT (spin transfer torque) switching mechanism; c) + d): SOT (spin orbit torque) switching mechanism; e) + f): VCMA (voltage-controlled magnetic anisotropy) effect. STT switching mechanism showing the orientation change of the FL caused by the interaction with passing electrons for (a) AP (anti-parallel) to P (parallel) state, and (b) P to AP state (PL: Pinned Layer; FL: Free Layer). (c) SOT-MRAM devices structure where an MTJ is placed on top of a heavy metal (SHE: Spin Hall Effect). (d) Schematic of the Spin Hall Effect, where the spin-up and spin-down electrons are separated to exert a spin-torque to switch the FL (HM: Heavy Metal; $J_S$: Spin current density; $J_C$: Charge current density). (e) VCMA-MTJ device structure ($V_b$: bias voltage). (f) Influence of different bias voltages ($V_b$) on the energy barrier $E_b$ of the FL are illustrated. $V_c$ is the critical voltage for which $E_b$ is removed.

magnetic anisotropy (VCMA) effect is used to assist the switching.

### 3.3.1. Spin-Transfer Torque (STT)

In STT-MRAM, a relatively high current density is passed directly through the MTJ. Depending on the direction of the current, spin-polarized electrons transfer angular momentum to the FL, exerting a torque on its magnetization. **Figure 5a,b** illustrates switching from AP to P and P to AP, respectively. A separate, lower-current operation can be used for reading the MTJ resistance while minimizing the probability of disturbing the FL magnetization.[25–28] A typical device-level R-V characteristic is shown in Figure 2g, illustrating the switching between P and AP states depending on the voltage polarity.

### 3.3.2. Spin–Orbit Torque (SOT)

In SOT-MRAM, the MTJ is typically placed on a non-magnetic heavy metal layer that exhibits strong spin-orbit coupling (SOC) (see Figure 5c).[29] The write operation is performed by applying a charge current through the heavy metal, generating a transverse spin current by the spin Hall effect (SHE)[30] which injects spins into the free layer that exerts torque on its magnetization (see Figure 5d).[31,32] This can lead to deterministic switching for in-plane FL magnetization,[30] while achieving deterministic switching in an out-of-plane magnetized FL requires breaking the in-plane symmetry of the device which can be achieved through a variety of methods.[33,34]

### 3.3.3. Voltage-Controlled Magnetic Anisotropy (VCMA)

The voltage-controlled magnetic anisotropy (VCMA) effect provides an electric-field-driven mechanism through which the

MTJs' state can be controlled. In CoFeB/MgO devices, the interfacial charge modulation, along with spin-orbit coupling explains the control of the interfacial perpendicular magnetic anisotropy[35–37] of the FL. When an electric field is applied, charges accumulate at the interface, resulting in a typically odd-dependence of the magnetic anisotropy on voltage, and consequently lowering or increase of the FL energy barrier (Figure 5e).[28,38] The VCMA effect is often used to assist SOT- and STT-based MTJ devices to lower the necessary switching current by reducing the energy barrier between the magnetization states (Figure 5f).[39] By applying a bias voltage $V_b$ at the top of a SOT device the energy barrier $E_b$ of the MTJ is lowered by the VCMA effect,[40,41] allowing the FL magnetization to switch more easily. This technique can also be combined with STT for improved switching efficiency, especially for smaller MTJ devices.[32] Similarly, the modulation of the energy barrier by VCMA can also be used in the absence of any STT or SOT torque, where it can drive the device toward precessional or thermally-activated switching for application as an entropy source.[42,43]

### 3.4. Alternative Electronic Materials

In addition to the previously discussed material systems, such as oxides, chalcogenides, and other compounds, recent research has highlighted the potential of 2D materials for advanced memristive and spintronic applications. Materials like hexagonal boron nitride (hBN)[44] and transition metal dichalcogenides (TMDs), including molybdenum disulfide ($MoS_2$), have enabled novel device architectures such as atomristors[45,46] and memtransistors[47]. In atomristors, hBN acts as the switching material, similar to that used in RRAM devices, but because it is crystalline, it has fewer defects, which helps reduce variability. Memtransistors are multi-terminal hybrid devices that integrate memristive behavior with transistor-like gate control, offering enhanced tunability. Furthermore, 2D materials are also being explored for spintronic applications, leveraging properties of 2D ferromagnetic materials such as $Fe_3GeTe_2$ (FGT).[48]

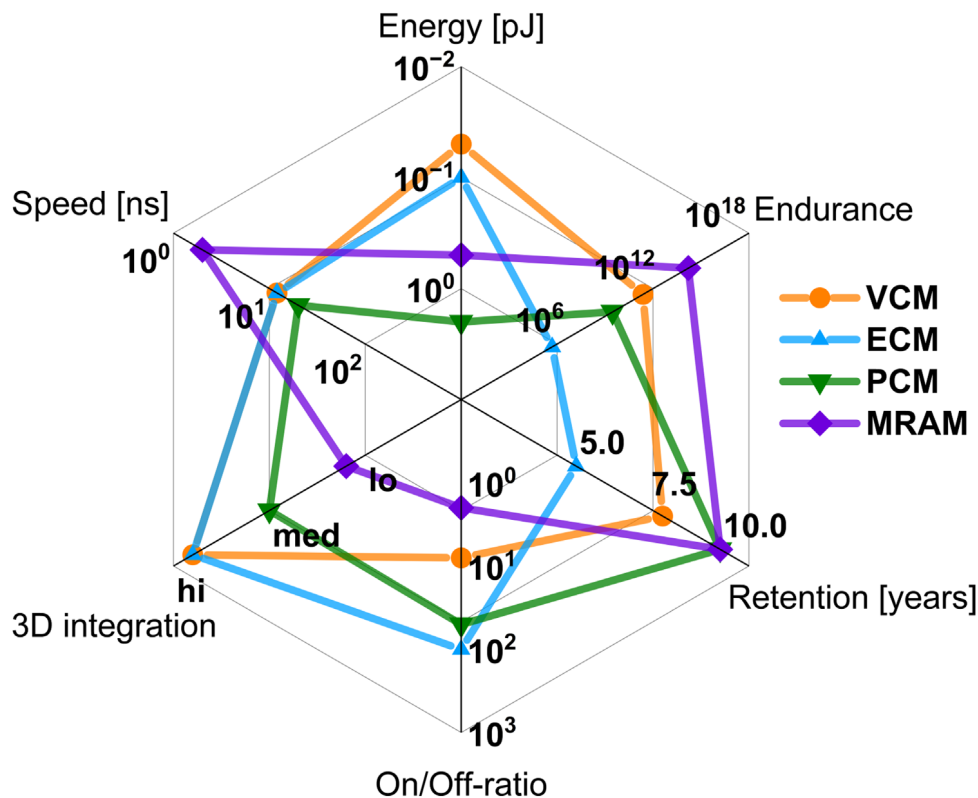

**Figure 6.** Comparison of the main device parameters of five different emerging memory technologies: VCM, ECM, PCM, MRAM (STT and SOT). The more area a graph covers the better the device parameters. Data was collected from refs. [49,139,142–144].

### 3.5. Comparison of the Device Characteristics

Comparisons among the main memristive technologies often focus on their endurance, retention, resistance ratio (for simplicity, we use instead of TMR for MRAM devices the resistance ratio for comparison with other technologies), CMOS compatibility, and switching speed. In our analysis, STT-MRAM and SOT-MRAM are grouped as MRAM devices because many of their core characteristics stem from the underlying MTJ structure.

**Figure 6** shows a radar chart of these technologies. RRAM devices (VCM and ECM) and PCM offer similar switching speed. RRAM outperforms PCM regarding energy consumption, but exhibits higher cycle-to-cycle and device-to-device variabilities.[10] MRAM devices excel in endurance and read/write speed, but are limited by their 3D integration capabilities and relatively low resistance ratios (on the order of 2 to 3 for P and AP states). By contrast, VCM cells offer resistance ratios of approximately an order of magnitude, and ECM and PCM devices often exceed two orders of magnitude. Retention of over 10 years is achieved by almost all device types (for ECM retention is a major drawback), and all devices are highly compatible with CMOS integration.[49] Overall, each device technology has unique trade-offs that influence its suitability for different memristive logic families and applications. Beyond comparing different memristive technologies among themselves, it is essential to consider their position relative to established memory solutions. To date, stand-alone memristive memory technologies are currently more expensive to manufacture than NAND Flash and DRAM. However, embedded memristive memories are emerging as a highly promising alternative[10] as their production cost is reported to be three to four times lower than other non-volatile technologies, such as embedded Flash. Furthermore, memristors offer superior scalability below 28 nm nodes, which remains challenging in embedded Flash due to high thermal budget and multiple lithography steps[49]. These advantages highlight the potential of memristors in modern and future commercial applications with competitive market prospects. To fully assess the potential of memristive technologies, it is essential to understand the underlying failure mechanisms that impact their reliability and performance.

## 4. Failure, Reliability, and Mitigation Analysis

This section examines the reliability challenges associated with memristive technologies, focusing on their underlying physical characteristics. The key aspects of reliability are as follows:

1. **Cycle-to-cycle (C2C) and device-to-device (D2D) variability**: C2C variability refers to fluctuations in the LRS and HRS of the same device across multiple switching cycles. D2D variability arises from the differences in switching behavior among different cells under identical programming conditions (voltage and pulse duration). In RRAM devices, C2C and D2D variability cannot be statistically separated.[50–52] The variations for RRAM and PCM devices in the programmed resistance state are stochastic due to imperfect fabrication processes with defects and the stochastic nature of the set and reset programming.[53] Especially in RRAM devices, the LRS varies due to variations in the shape of the filament and defect concentrations in the cell when its created.[54,55] PCM variability is influenced by the shape and size of the crystalline grains in the phase change material. Additionally, a Gaussian spread in the activation energy for crystallization, along with variations in the active amorphous volume, result in variations of the crystallization time and SET resistance states. These variations contribute to both C2C and D2D variability.[56] In addition to material engineering, programming techniques such as multi-pulse schemes can mitigate these variabilities.[11] Although less studied, fabrication differences among MRAM devices drive D2D variability. Most importantly, while remarkable cross-wafer thickness uniformity is possible with state of the art MRAM film deposition tools used in industry, the etch process is more complicated due to the presence of many dissimilar materials in the stack and the ultra-thin tunnel barrier. Hence, an un-optimized etch can be a source of significant D2D resistance variation. The C2C variability of MRAM is predominantly due to thermal fluctuations[57] and the intrinsically stochastic nature of the spin-based switching of STT.[26,58]

2. **Retention failure**: Retention refers to a device's ability to retain its data state over time.[59] In RRAM devices, the LRS shows better retention than the HRS.[50] In PCM devices, the amorphous HRS is metastable, making it susceptible to undesired crystallization which degrades retention. Additionally, crystallization variability also affects the retention of PCM devices.[60] In MRAM devices, retention failures arises from thermal instability of the free layer magnetization,[61,62] which can be improved by increasing the thermal stability factor (i.e., FL anisotropy and volume).[63]

3. **Endurance failure**: Endurance is the ability of a device to undergo repeated reliable SET/RESET switching cycles without significant degradation or failure.[59,64] Especially in RRAM devices, repeated switching can gradually diminish the resistance ratio until the device becomes stuck in one state.[64] Endurance enhancement can be achieved through various mitigation techniques, such as device-level engineering[65] or applying triangular pulses to decrease the current overshoot during switching.[66–68] From a circuit level perspective, memory lifetime can be improved by adding spare parts for failed devices or by predicting and reinforcing weak cells before failure through dynamic spare reallocation.[69] Another approach is wear leveling—redistributing write operations evenly across memory cells.[70] PCM endurance is largely constrained by high temperatures, current density, and electromigration, particularly at the interface between the heater and the programmable region.[71] MRAM devices can achieve high endurance ($1 \times 10^{15}$ for MRAM, while PCM achieves $\approx 1 \times 10^9$, and RRAM between $1 \times 10^6$ and $1 \times 10^{12}$), but the repeated application of large switching currents in STT devices may lead to oxide barrier breakdown.[72] In SOT-MRAM diffusion mechanisms are causing degradation of MTJs during

SOT stress.[73] In particular, optimizing pulse amplitude and pulse duration reduces electrical stress and enhances device endurance.[74]

4. **Write failure/disturbance**: Write failures, also referred to as write error rate (WER) in the context of MRAM devices, occur when a cell fails to switch to the intended state during programming. In contrast, write disturbance refers to the unintentional switching of neighboring cells.[64] In RRAM devices, a program-verify approach can be employed to reduce write failures, which also enables programming with a tighter resistance distribution.[75] Another key concern are sneak path currents during writing, often alleviated by adding a selector device. Neighboring cells in PCM can also be thermally disturbed during write operations.[71] In STT-MRAM and SOT-MRAM, which both contain an MTJ, the stochastic nature of spin-based switching contributes to write failures.[61,76] Perpendicular MTJs generally exhibit lower WERs.[77] Further mitigation techniques for STT-MRAM are scaling down the MTJ as the sensitivity to parameter variations on WER is reduced.

5. **Read instability/read noise/RTN (random telegraph noise)**: When the resistance value of a memristive device is read multiple times, fluctuations in the read current may occur without changing the stored state. In RRAM, read noise in HRS can reach up to 10%,[55] explained by random oxygen-vacancy jumps.[50] In PCM, spontaneous transitions between the crystalline and amorphous phases may cause read disturbances.[71] Due to the nanometer-scale dimensions of RRAM filaments and PCM conductive paths, the trapping and de-trapping of single electrons can lead to detectable read noise. In STT-MRAM, read disturbances occur when the read current unintentionally flips the free layer magnetization. Additionally, a false read-out can happen due to read time failures, where the read time is too short for reading the cell state. Optimizing process parameters, geometry, and materials can help mitigate such disturbances.[63,77] The VCMA effect can be used to stabilize the FL during readout, which has been shown to eliminate read disturbance in MRAM bits.[78,79] Again, sneak path currents are also critical for reliable resistance reading.

Each memristive technology faces distinct reliability challenges tied to its switching mechanism. For example, C2C and D2D variabilities are particularly problematic for RRAM, whereas MRAM faces performance challenges primarily from read and write disturbances.[10] These reliability issues can degrade device performance, leading to erroneous logic outputs and potentially inoperable logic gates. It is therefore crucial to implement appropriate measurement schemes and device-level optimizations to identify and address these challenges.[80] Most of the failure and variability mechanism explained are considered as soft errors and allow to be corrected by error correction code (ECC) and redundancy (like triple modular redundancy) to improve reliability.[81–83] ECC is typically used to correct indirect errors caused by gradual state drift, while triple modular redundancy addresses direct errors caused by incorrect operations by replicating the same logic operations. The following section provides a detailed overview of the most well-known stateful and non-stateful logic techniques. It includes their functionality, existing simulation and experimental results, as well as the suitability and reliability of each device technology for the corresponding logic technique.

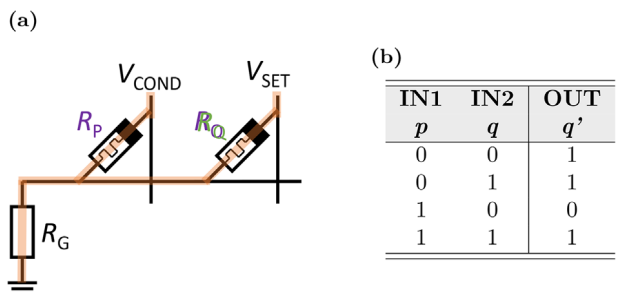

**Figure 7.** IMPLY with a) RRAM devices[84] (replaceable by PCM[91] and MTJ) devices[92]) and b) IMPLY truth table. The purple annotations show the logic input variables while the green annotations depict the logic output variable. The current path is highlighted in orange. One of the input memristors, $Q$, is also the output memristor.

## 5. Logic Families

This section surveys various memristive logic families and their underlying principles, focusing on two primary categories: stateful logic and non-stateful logic. In stateful logic, inputs and outputs are stored in memristive devices that retain their resistive states during computation. In contrast, non-stateful logic employs different physical representations for inputs, outputs, or both, often requiring additional read/write steps.

### 5.1. Stateful Logic

#### 5.1.1. IMPLY

One of the earliest stateful logic techniques is material implication logic, also called IMPLY or IMP, introduced by Borghetti et al.[84] The schematic of an IMPLY gate is shown in **Figure 7a** and comprises of two memristors, $P$ and $Q$, placed in parallel and connected to ground through a series resistor $R_G$. The inputs are represented by the resistive states $p$ and $q$ of the two memristive cells $P$ and $Q$, respectively, while the output $q'$, is given by the resistance of $Q$ after the operation. Figure 7b shows the corresponding truth table.

During the IMPLY operation, two different voltages, $V_{SET}$ and $V_{COND}$, are applied to cells $Q$ and $P$. The higher voltage, $V_{SET}$, can induce a SET transition, whereas $V_{COND}$ is deliberately kept too low to switch any device by its own. Applying both voltages simultaneously creates a voltage divider that either triggers a SET in $Q$ if $P$ is in the HRS or leaves $Q$ unchanged if $P$ is in the LRS.

Despite its promise of fast, low-energy computing in memory, IMPLY suffers from several drawbacks. First, different input combinations require different relative ratios of $V_{SET}$ and $V_{COND}$, making it difficult to select a single pair of voltages that is robust to variations.[85] Another issue is that repeated application of voltages can inadvertently alter the resistive states of the input memristors, causing instability.[86]

An enhanced design, SIMPLY, was proposed by Puglisi et al.[87] SIMPLY relies on the fact that in the IMPLY function,

the only case where $q' \neq q$ is when $p = q = 0$. Hence, the memristor must only perform a SET process in this case. In SIMPLY, a comparator first detects in a read-step if $p = q = 0$. If so, a SET voltage is applied to $Q$. This read-then-write approach eliminates state degradation and narrows the voltage requirements, but at the cost of extra peripheral components and increased latency.

Most IMPLY experimental demonstrations have used VCM-type RRAM devices,[88,89] though implementations with ECM,[90] PCM,[91] and STT-MRAM[92] have also been reported. Jin et al. have shown the IMPLY function as well as NOT, OR, and COPY with ECM devices, including the implementation as a full adder.[90] Hoffer et al. demonstrated IMPLY with PCM devices similar to RRAM devices, with a chosen value of $V_{COND}$ equal to $V_{SET}/2$.[91] Cao et al.[92] experimentally demonstrated OR, IMP, NAND, and NIMP(¬IMP) stateful logic functions based on the IMPLY design with MTJs based on STT by varying the voltages $V_{COND}$ and $V_{SET}$.

A key challenge in IMPLY logic lies in choosing voltages that ensure reliable switching without disturbing input states, especially in devices (e.g., RRAM) with higher variability. Consequently, emerging memory technologies like PCM and MRAM, which typically exhibit lower variability, may prove more suitable for dependable IMPLY-based operations.

#### 5.1.2. Memristor Aided Logic (MAGIC)

MAGIC is a prominent stateful logic technique that improves upon IMPLY in terms of area, latency, and energy.[93]

In contrast to the IMPLY function, where the output is written to one of the two input cells, the MAGIC logic operation stores the output in a third cell. This approach preserves the states of the input memristors. The logic gate of a MAGIC NOR gate is presented in **Figure 8a**. A voltage $V_G$ is applied at two parallel input cells, both connected in series to a third output cell that is grounded. Due to the connection configuration, the polarity of input and output cell are opposite, so that $V_G$ can only induce a RESET of the output cell (see definition in [94]).

The resistance of the output is initialized to be low, which represents a logical '1' and the voltage across the output memristor depends on the input resistance values. If the voltage over the output cell exceeds its RESET voltage, the output memristor resets to logical '0', otherwise, it remains unchanged. For example, if both input resistances are high, the majority of the voltage drops across the input cells. As a result, only a small voltage drop appears across the low-resistance output cell, which is insufficient to induce a RESET switching process. Therefore, the output remains '1' for both inputs being in the HRS (logical '0'). For all other cases, at least one input is '1' (low resistance value), leading to a voltage drop over the output, which is higher than the RESET voltage. In these cases, the output switches its state to '0'. This functionality describes the NOR function.

The use of a separate cell for the output and the preservation of the input states enable cascading of operations without initialization.[94]

MAGIC was introduced with five basic gates—NOR, NAND, OR, AND, and NOT—but only the MAGIC NOR gate is crossbar compatible.[94] Additional MAGIC gates have been proposed later on,[95–97] out of which some are also crossbar compatible.

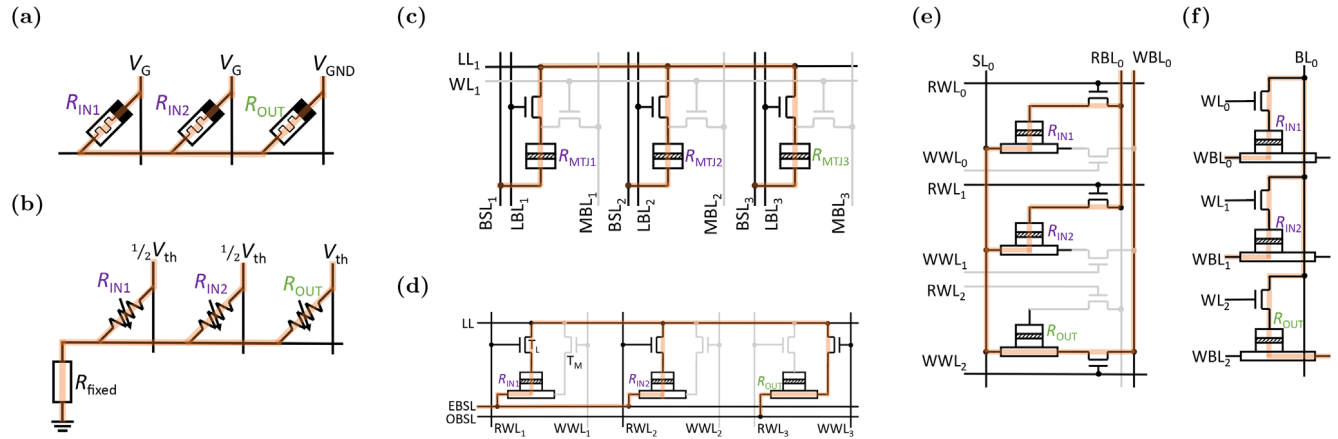

**Figure 8.** MAGIC logic gates with their variations corresponding to the requirements of different devices: a) Original gate design for VCM/ECM devices.[97] b) Gate for PCM devices.[91] c) CRAM with STT-MRAM devices (STT-CRAM)[102] (acronyms: LL = Logic Line, WL = Word Line, BSL = Bit Select Line, LBL = Logic Bit Line, MBL = Memory Bit Line). d) CRAM with SOT-MRAM devices (SpinPM[104] derived out of SHE-CRAM[145]) (acronyms: EBSL = Even BSL, OBSL = Odd BSL, RWL = Read Word Line, WWL = Write Word Line). e) Gate for SOT-MRAM devices with a standard 2T1MTJ array structure[105] (acronyms: SL = Select Line, RBL = Read Bit Line, WBL = Write Bit Line). f) VGSOT-MRAM array that is assisted by VCMA[105] (acronyms: WL = Word Line). The purple annotations show the logic input variables while the green annotations depict the logic output variable. The current path is highlighted in orange.

This general explanation of a MAGIC gate is applicable to its application using different emerging memory technologies. Therefore, in the following subsections we show the implementation of MAGIC gates with different devices and crossbar arrays.

### 5.1.3. MAGIC with RRAM

MAGIC has been experimentally demonstrated with VCM devices.[97,98] In VCM-based implementations, the MAGIC NOR gate is hampered by input instability as the required SET voltage must be at least twice the RESET voltage. This derives from the voltage divider shown in Equation (1).

$$V_{OUT}(R_{IN1}, R_{IN2}) = V_G \cdot \frac{R_{ON}}{R_{IN1} || R_{IN2} + R_{ON}} \qquad (1)$$

With $R_{OFF} >> R_{ON}$, $V_G$ must be greater or equal to $2|V_{RESET}|$ in order to switch the output in all cases where at least one input cell is in the LRS. However, the input cells are configured such that a SET process can take place if the applied voltage is higher than the required SET voltage. A limitation of this approach is that for RRAM devices $V_{SET} < |V_{RESET}|$, so that the SET voltage is always exceeded if $V_G \geq 2|V_{RESET}|$. Therefore, the inputs can change their state during the operation, leading to input instability. This may result in both inputs switching to the LRS, which alters the logical output and invalidates the NOR operation.

As a result, adaptations to the input voltages and initial resistance values have been proposed to realize OR, NIMP, and XOR gates.[97] A recent study investigated the implementations of OR and NOT gates within a 1T1R (1 transistor 1 resistor/memristor) crossbar array using a VCM device, with particular focus on energy distribution across different gate operations. Notably, the initialization process accounts for the majority of energy consumption.[98]

A dedicated failure mechanism analysis framework for RRAM-based MAGIC further assesses various failure modes tied to log-ical input and outputs across different crossbar sizes.[99] This framework additionally contrasts Scouting Logic (detailed in Section 5.2.1) with MAGIC, highlighting their respective operational and reliability considerations.

### 5.1.4. MAGIC with PCM

MAGIC NOR gates have been experimentally demonstrated using PCM devices.[91] Because PCM exhibits a unipolar switching mechanism, a modified circuit design is required (see Figure 8b). For instance, OR and NIMP gates can be implemented with only three cells, where input and output memristors are kept separate. In the case of the NOR gate, a grounded resistor is used to connect to the logic memristors - similar to the IMPLY configuration - and can be easily integrated into the peripheral circuit. The successful experimental and statistical evaluations are encouraging and pave the way for further research into digital PIM architectures.

### 5.1.5. MAGIC with STT-MRAM

Early demonstrations of MAGIC gates were carried out in simulations with STT-MRAM devices.[100] In this approach, a 1T1MTJ array (1 transistor 1 MTJ) was designed to perform MAGIC NOR operations. However, standard arrays lack a shared bitline for input and output memristors along a row, posing a key challenge. Because MTJs are current-controlled, the necessary switching voltage must be derived from each MTJ's resistance and threshold current. Selecting a transistor gate voltage common to all transistors in the array adds complexity. Furthermore, Monte Carlo simulations indicate that the error rate of a wrong logic operation must be reduced to obtain a more robust performance.

A similar architecture, called computational RAM (CRAM[101]), introduces further array designs for spin-based stateful logic with additional logic lines (LL) and an extra transistor as in Figure 8c. One transistor handles write/read data operations, while the second transistor enables logic via the LL. Although this approach uses current-steered logic and has higher area overhead than 1T1MTJ, simulations suggest it alleviates key parasitic issues in large areas, such as reducing bit select line (BSL) parasitics.[102] Recently, the STT-CRAM concept was experimentally demonstrated in a 1x7 array for NAND gates.[103]

### 5.1.6. MAGIC with SOT-MRAM

SOT-MRAM features three-terminal cells with dedicated read and write paths. Several designs for implementing MAGIC on SOT-MRAM arrays have been proposed.[104,105]

In the approach of Hoffer et al.,[105] a 2T1MTJ standard memory array is utilized, as shown in Figure 8e. For logic operations, the read transistors of the input cells and the write transistor of the output cell are opened. The current through the heavy metal layer of the output transistor depends on the resistance of the input cells, and if it exceeds a critical switching current, the output MTJ toggles its state. However, the inherently small resistance ratio between the two MTJ states leads to a reduced read margin, which is further affected by device-to-device (D2D) variations. As a result, distinguishing between high and low resistance states across different cells becomes increasingly error-prone. Moreover, since this logic mechanism relies on voltage dividers, the resistance ratio of the device determines the logical output. A lower resistance ratio reduces reliability. Additionally, achieving a sufficiently high accumulated read current, higher than the critical current for SOT switching, can reduce array density as it increases the size of the MTJs. Thermal noise and process variations also impact the error rate.

A notable example of stateful SOT-MRAM logic is *SpinPM*,[104] which adapts a CRAM-like layout, as shown in Figure 8d. Here, a voltage $V_0$ is applied at the EBSL (Even Bit Select Line), while the input rows and output write wordline are enabled, creating a current path through the input cells and the SHE (Spin Hall Effect) channel of the output. If the combined current exceeds the critical threshold, the output MTJ switches. Although experimental validations of SpinPM are pending, a related design, called SHE-CRAM has been fabricated, showing energy gains and faster switching relative to STT-CRAM, but at the cost of increased area.[106]

Other SOT-MRAM crossbar architectures, such as those using voltage-gated SOT (VG-SOT), exploit the VCMA effect to lower the write current for MTJ switching, as shown in Figure 8f.[105,107] Hoffer et al.[105] proposed grounding the output bitline and applying a voltage to WL0. Then, sufficient current passes through the shared bitline only if the input memristors' combined resistance is sufficiently low. The VCMA effect significantly decreases the critical current required. However, row-wise MAGIC operations are not feasible; instead, column-wise logic is used. The SOT channel (heavy metal) is shared between rows and allows denser arrays since the switching is only enabled through the VCMA effect when a voltage is applied at the bitlines. Simulation results highlight challenges related to variations in the critical switching current, which can cause logic failures if the output MTJ does not switch as intended.

### 5.2. Non-Stateful - Near-Memory Logic

#### 5.2.1. Pinatubo and Scouting Logic

A promising candidate for non-stateful near-memory logic is Pinatubo[108] - Processing In Non-volatile memory ArchiTecture for bUlk Bitwise Operations. In this approach, the logic computation is shifted from the memory (in-memory-computing) to a redesigned read circuitry (near-memory) which is able to compute a bitwise logic of two or more memory rows in an efficient manner. The integration on circuit level is explained by using the word *Scouting Logic*.

Xie et al.[109] introduced the main idea of implementing the Pinatubo logic technique on a circuit level, namely *Scouting Logic* (SL).[109] We use the word *Scouting Logic* when referring to this logic concept, as this paper focuses on circuit level rather than architecture level implementations (such as Pinatubo). In SL, the inputs are stored in the resistive states of two memristors. The output is determined by measuring the combined current flowing through these memristors, as shown in **Figure 9a**. This measured current is compared against a reference current $I_{ref}$, such that currents lower or higher than $I_{ref}$ correspond to logical '0' or '1', respectively. The value of the reference current is determined by the desired logic operation. Figure 9b illustrates the SL configuration, including the reference current placement for AND, OR, and XOR operations. Notably, the input combination '10' (cell 1: LRS, cell 2: HRS) and '01' (cell 1: HRS, cell 2: LRS) yield the same current; thus, they produce identical outputs. To implement the AND function, $I_{ref}$ is set between the currents corresponding to '10'/'01' and '11', such that the output is '1' only when both inputs are '1'. For the XOR function, two distinct reference currents are needed because the input combinations '00' and '11' produce the same output ('0').

A key advantage of SL is its broad compatibility with existing array structures. It can also be extended to more than two cells, as long as the different states remain clearly distinguishable. However, SL relies on accurate sense amplifiers and low device variability; a particular challenge with RRAM device due to their variability. Although SL can be adapted to different memristive technologies, each comes with distinct device specific challenges.

#### 5.2.2. Enhanced Scouting Logic

An enhanced version, *Enhanced Scouting Logic*, was proposed by Yu et al.,[110] employing a 2T1R array configuration. By selectively connecting memristors in series or parallel, the approach simplifies reference current placement for AND and OR operations, although it does not support XOR. Figure 9c shows the OR gate operation.

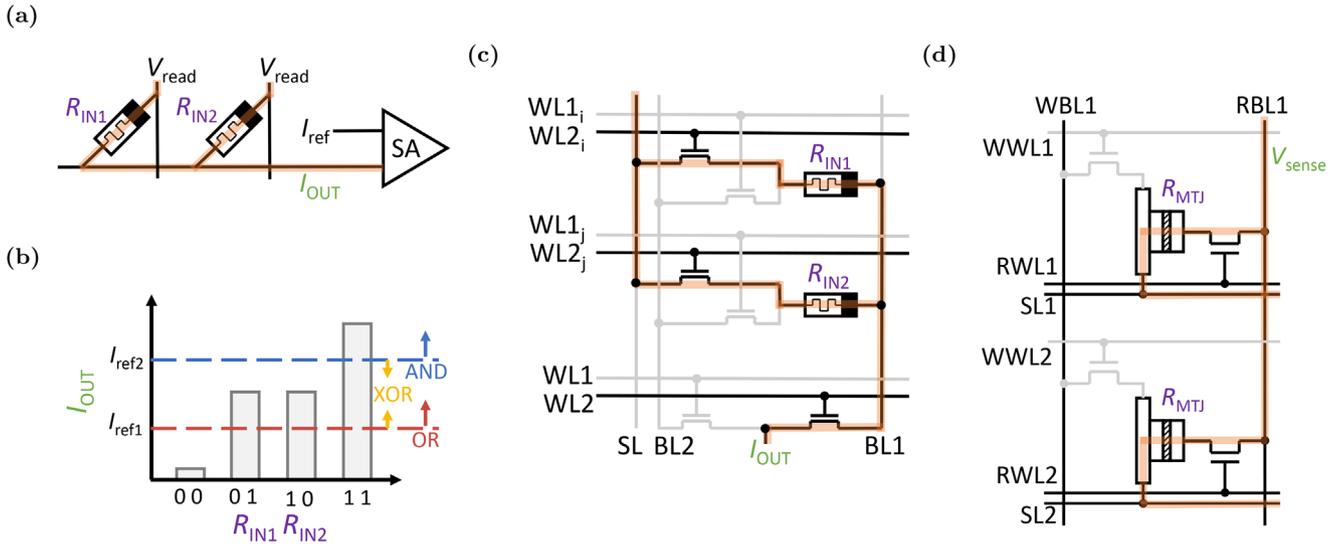

**Figure 9.** Non-stateful near-memory logic techniques with a) Scouting Logic with RRAM,[109] b) Scouting Logic functionality of AND, OR and XOR logic gate, c) OR gate with Enhanced Scouting Logic[110] (acronyms: WL = Word Line, SL = Select Line, BL = Bit Line), d) PIMA-logic with SOT-MRAM[118] (acronyms: WBL = Write Bit Line, RBL = Read Bit Line, WWL = Write Word Line, RWL = Read Word Line). The purple annotations show the logic input variables while the green annotations depict the logic output variable. The current path is highlighted in orange.

### 5.2.3. Scouting Logic with RRAM

To date, most experimental demonstrations have used VCM devices[111–113] which suffer from relatively high variability and a relatively small resistance ratio, risking overlap between logical states. Nonetheless, ECM devices—with their potentially high resistance ratios—may be particularly promising for SL, though additional research is needed to confirm feasibility.

### 5.2.4. Scouting Logic with PCM

For the implementation of SL, PCM devices are proposed and their functionality is confirmed experimentally by[114,115]. It showed good suitability of PCM devices due to their high resistance ratio, which offers a reliable state differentiation.

### 5.2.5. Scouting Logic with MRAM

Until now there are no experimental demonstrations using MRAM devices in the standard SL design. However, the smaller tunnel magneto-resistance ratio (TMR) in magnetic tunnel junctions (MTJs) further narrows the sensing margin (Figure 9b), especially if multiple rows are activated simultaneously or under temperature/process variations.[63,116] Using different cell structures, HieIM (Highly Flexible InMemory) computing[117] and PIMA (Processing-in-Memory Architecture)-logic[118] are implemented. A 1T1R structure is used in the HieIM with STT-MRAM devices, while PIMA targets SOT-MRAM technology. During a logic operation, RWL1 (Read Word Line), RWL2, and RBL1 (Read Bit Line - connected to the SA) are activated, while SL1 and SL2 are grounded (see Figure 9d). The resulting current generates a sense voltage $V_{sense}$ in voltage-mode SA, enabling AND/NAND and OR/NOR gates. As in other near-memory schemes, the SA is a crucial component of PIMA-logic, underscoring its central role in translating device currents into digital currents. This is especially relevant in MRAM due to the low resistance ratio and D2D variations of the resistance states.

### 5.3. Non-Stateful - In-Memory Logic

### 5.3.1. Functionally Complete Logic with 1T1R

Applying voltages to multiple terminals in a 1T1R (1 Transistor 1 Resistor) memory array can enable non-stateful in-memory logic. Given a memristor and a transistor, as shown in **Figure 10a**, all 16 Boolean functions can be realized by the configurations given in Figure 10b.[119] The functions compute the output of two inputs (p and q) by defining the configuration of the transistor gate (G), the top electrode (TE = T1), and the source of the transistor (equals to the bottom electrode (BE = T2) for an open transistor), and the memristor's initial resistance state (I = R). The output is the resistance state of the memristor after the operation, which can be read out by applying a read voltage while the transistor is open.

As an example, the AND function changes the initial resistance state of '0' only when the operation voltage is applied at the top electrode (denoted as q) and the transistor is open (denoted as p). This condition is satisfied only for both inputs p and q being 1. In all other cases, the memristor stays constant at 0. This functionality therefore, represents the AND function.

If a cell fails to switch when expected (writing failure) or unintentionally changes its state (retention failure) the output can become corrupted. Furthermore, a clear distinction between HRS and LRS is crucial, requiring both low variability and a high resistance ratio. ECM devices typically offer higher resistance ratio than VCM devices, but at the cost of lower endurance. This logic techniques relies on the bipolar switching behavior of the

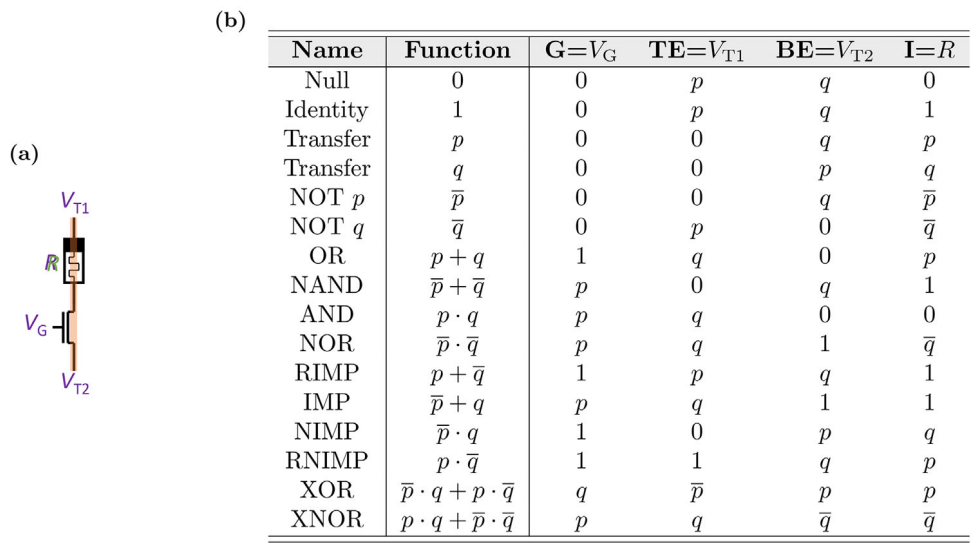

(a)

$V_{T1}$

$R$

$V_G$

$V_{T2}$

(b)

| Name | Function | $G=V_G$ | $TE=V_{T1}$ | $BE=V_{T2}$ | $I=R$ |
|---|---|---|---|---|---|
| Null | 0 | 0 | $p$ | $q$ | 0 |
| Identity | 1 | 0 | $p$ | $q$ | 1 |
| Transfer | $p$ | 0 | 0 | $q$ | $p$ |
| Transfer | $q$ | 0 | 0 | $p$ | $q$ |
| NOT $p$ | $\bar{p}$ | 0 | 0 | $q$ | $\bar{p}$ |
| NOT $q$ | $\bar{q}$ | 0 | $p$ | 0 | $\bar{q}$ |
| OR | $p+q$ | 1 | $q$ | 0 | $p$ |
| NAND | $\bar{p}+\bar{q}$ | 0 | 0 | $q$ | 1 |
| AND | $p\cdot q$ | $p$ | $q$ | 0 | 0 |
| NOR | $\bar{p}\cdot\bar{q}$ | $p$ | $q$ | 1 | $\bar{q}$ |
| RIMP | $p+\bar{q}$ | 1 | $p$ | $q$ | 1 |
| IMP | $\bar{p}+q$ | $p$ | $q$ | 1 | 1 |
| NIMP | $p\cdot\bar{q}$ | 1 | 0 | $p$ | $q$ |
| RNIMP | $\bar{p}\cdot q$ | 1 | 1 | $q$ | $p$ |
| XOR | $\bar{p}\cdot q+p\cdot\bar{q}$ | $q$ | $\bar{p}$ | $p$ | $p$ |
| XNOR | $p\cdot q+\bar{p}\cdot\bar{q}$ | $p$ | $q$ | $\bar{q}$ | $\bar{q}$ |

**Figure 10.** Non-stateful in-memory 1T1R logic technique with a) 1T1R logic circuit with RRAM device,[119] b) 1T1R logic table with all possible gates that can be implemented. The purple annotations in (a) show the logic input variables while the green annotations depict the logic output variable. The current path is highlighted in orange.

memristor cell. Therefore, the realization with unipolar devices such as PCM would require an adjustment from polarity-based switching to amplitude-based control. STT-MRAM devices face challenges with their relatively small resistance ratio, which can lead to overlapping HRS and LRS values or read disturbances in the read-out step, leading to an incorrect output. SOT-MRAM devices do not have read disturbance issues due to separate read/write paths,[120] but they deviate from a 1T1R structure, making the existing logic table inapplicable. Overall, further investigation is needed for harnessing PCM and MRAM devices for this logic scheme.

### 5.3.2. Complementary Resistive Switch (CRS)

CRS[121,122] was introduced to mitigate sneak-path currents in memory applications by using two bipolar memristors connected in series with opposite polarities, as shown in **Figure 11**a. Alternatively, CRS behavior can be obtained by a single-stack memristor capable of *complementary switching* (CS) which replicates the same behavior.[123]

Figure 11c shows a schematic I-V characteristic for a CRS cell. The two memristors, A and B, can store a logical "0" if A = HRS and B = LRS, or a logical "1" if A = LRS, B = HRS. To determine the logical output, an I-V sweep is performed. An I-V sweep reveals the stored state by observing current spikes and RESET transitions at defined thresholds. For instance, if the CRS was in logical "0", applying a positive voltage above $V_{th,1}$ would switch memristor A to LRS, leading to a spike current. Then, a higher voltage above $V_{th,2}$ would reset memristor B to HRS, changing the logical state of the CRS from "0" to "1". If the CRS was in logical "1", the resistances of the memristors would be unchanged, avoiding current spikes.

This CRS characteristic can be used for logic operations, such as AND (see Figure 11b), through a particular sequence of applied voltages. The voltages applied at T1 and T2 represent the inputs (named as $p$ or $q$) and the resistance $R_{CRS}$ represents the output. Read-out is performed through spike detection by applying a positive voltage sweep. Since the read-out is destructive, an inverse (negative) read-out pulse is needed to restore the original state. Fourteen of the sixteen basic Boolean functions (mentioned in Figure 10b) can be implemented with a standard positive read-out ("10"), while the remaining two (XOR and XNOR) require an alternative read-out scheme ("01").[124]

Although simulation and experimental studies indicate CRS feasibility,[125] only a few have demonstrated actual logic operations experimentally.[126–128] Simon et al.[126] demonstrated logic operations with CRS experimentally for 14 logic gates with VCM and ECM devices. Further experimental results and device characteristics for CRS/CS are listed in **Table 1**, demonstrating the feasibility of CRS with either *I*–*V* curves or pulses. Most reported work relies on RRAM devices, primarily VCM, because CRS requires bipolar switching. PCM devices are generally unsuitable due to unipolar switching, and MRAM—with its transistor-based approach—does not benefit from CRS as CRS mainly addresses sneak path issues.

Moreover, the destructive read-out increases power consumption and reduces speed relative to non-destructive logic. While certain studies[129] propose non-destructive read-outs and margin-improving strategies,[130] CRS remains a challenging, but promising option for in-memory logic.

### 5.3.3. Complementary Switching with VCM

Practical hurdles include the high voltages needed for CRS switching, device variability, and the potential for polarity-independent resets. However, complementary switching (CS) devices combine the functionality of two stacked memristors, showing two switching events at each voltage polarity, in one device. They are only demonstrated in VCM and can reduce fabrication complexity, have longer lifetime, and perform computation with a lower switching voltage of 1.2 V. However, in CS devices, reliability is an issue as they should follow an *I–V* curve similar to Figure 11a, showing the intended four switching events per cycle. Instead, for some CS devices, only one of the two intended switching events takes place for each polarity. This behavior turns the CS device into a regular bipolar resistive switching device, leading to a failure of the ECM.[123,126] ECM devices exhibit a higher resistance ratio compared to VCM (Figure 6). However, they suffer from greater switching variability, including unintended polarity-independent RESET switching.

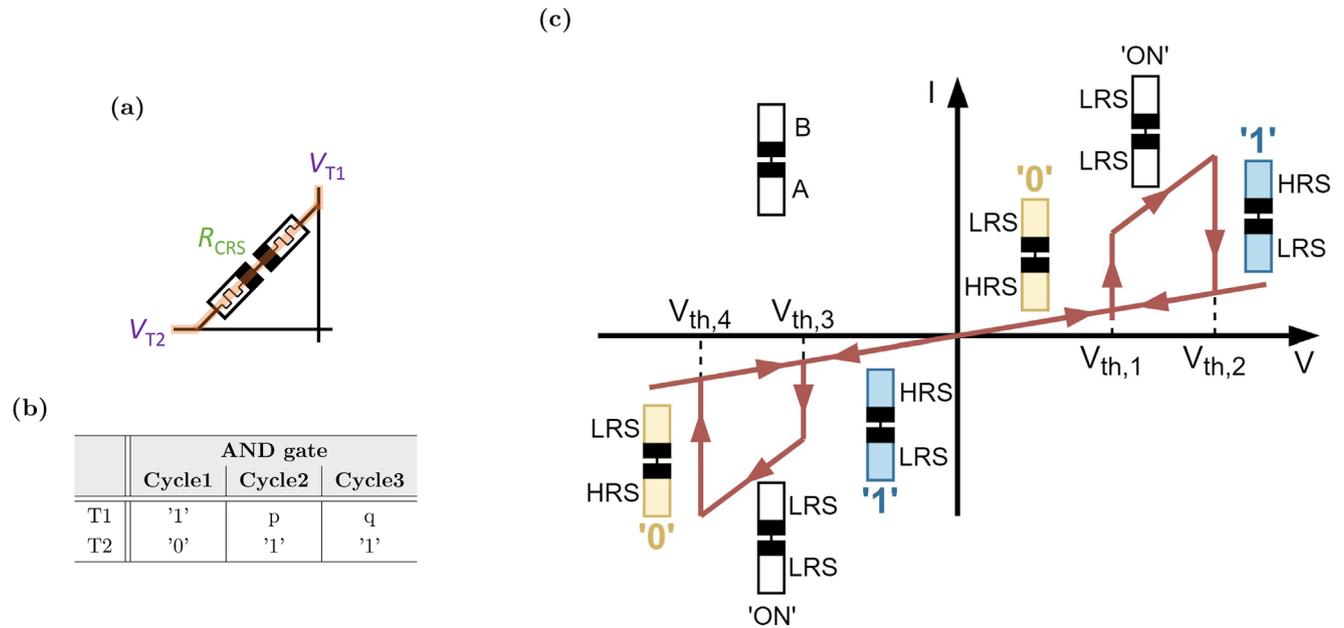

**Figure 11.** Non-stateful in-memory CRS logic technique with a) Complementary Resistive Switching (CRS) principle,[122] b) CRS AND gate logic table with $p$ and $q$ as input variables, and c) CRS I-V curve. A positive sweep represents a positive voltage '1' at memristor B (or $V_{T1}$) and A (or $V_{T2}$) is grounded '0'. For the negative sweep the voltages applied are reversed. The purple annotations in (a) show the logic input variables while the green annotations depict the logic output variable. The current path is highlighted in orange.

**Table 1.** Material characteristics regarding their logic family.

| Logic | Device | Stack | R-ratio | Endurance | $V_{SET}$ [V] | $V_{RESET}$ [V] | Pulse width | Power consumption | Refs. |
|---|---|---|---|---|---|---|---|---|---|
| MAGIC | VCM | Pt/Ta$_2$O$_5$/W/Pt | >10 | – | 1 | 2 | >500 µs | – | [97] |
| | VCM | Pt/TaO$_x$/W/Pt | 10 | – | 2 | 1.8 | 4 ms | 1300 nJ(set), 312 nJ(reset) | [98] |
| | STT | multi-layer stack of CoFeB/MgO | TMR 100% | – | 0.35 | 0.35 | 1 ms | – | [103] |
| IMPLY | VCM | Pt/ZrO$_2$/Ta/Pt | 20 | – | 1.1 | – | 100 µs | – | [89] |
| | ECM | Cu/GeTe/TiN | 10$^4$ | 10$^4$ | 1.4 | 1.6 | 60 ns | 58.5 pJ (set), 8.67 pJ (reset) | [90] |
| | ECM | Cu/Al$_2$O$_3$/Poly-Si | 1000 | – | 4 | −2 | 60 ms | 30 pJ | [152] |
| | STT | MgO/CoFeB/ W/CoFeB/MgO | TMR 100% | – | 0.7 | −0.7 | 1 ns | – | [92] |
| | STT | multi-layer stack [153] | TMR 90% | – | ∼ 0.5 | ∼ -0.7 | 3 ns | 4.5 pJ | [153] |
| Further Stateful Logic | ECM | Cu/α-Si/α-C/Pt | 1000 | 4·10$^8$ | 3.5 | 2 | 100 ns | – | [148] |
| | PCM | TiN/GST/SiO$_x$ | 10 | 10$^4$ | 1.2 | 3 | 100 ns | 0.5 mW | [105] |
| Scouting | VCM | TiN/HfO$_2$/Ti/ TiN | 20 | 10$^4$ | – | 3.25 | – | – | [112] |
| | VCM | x/HfO$_2$/x | – | 10$^6$ | – | – | – | – | [111] |
| | PCM | – | > 50 | – | 2.1 | 3.6 | 100 ns | 630 µW | [114] |
| 1T1R | VCM | x/HfO$_2$/x | >10 | 10$^7$ | 1.5 | 1.5 | 200 ns | – | [119] |
| | VCM | TiN/SiO$_2$/Ti | 20 | – | 1.3 | 1.6 | 1µs | – | [113] |
| CRS | VCM | Pt/TaO$_x$/Ta/ TaO$_x$/Pt | >10 | 1.25·10$^6$ | 1.9 | 1.9 | 500 ns | – | [126, 127] |
| | ECM | Pt/Ag/GeS$_x$/ Pt/GeS$_x$/Ag/Pt | 2·10$^4$ | – | 5 | 5 | 100 ns | – | [126, 154] |
| | ECM | Cu/SiO$_2$/Pt | >1500 | – | 5 | 5 | 200 µs | – | [155] |
| | VCM | Pt/TiO$_x$/TiO$_2$/ TiO$_x$/Pt | <10 | 10$^5$ | 1.2 | 1.2 | 5 ms | – | [156] |
| | VCM | Pd/TaO$_x$/Ta/Pd (CS) | 2 | 120 | 2 | 2 | 5 ms | – | [157] |
| | VCM | Pt/HfO$_2$/Hf/Pt (CS) | – | 10$^9$ | 2 | 2 | 500 ns | – | [128] |
| | VCM | TiN/HfO$_2$/TiN (CS) | 10 | 10$^3$ | 1.2 | 1.2 | – | – | [123] |
| V-R with PCM | PCM | TiN/GeSbTe/TiN | – | – | 2.2 | – | 300 ns | – | [132] |
| | PCM | TiW/Ge$_2$Sb$_2$Te$_5$/ TiW | >25 | – | 0.8 | 2 | 200 ns/ 30 ns | 0.57 pJ/ 14.17 pJ | [149] |
| | PCM | TiW/Ge$_2$Sb$_2$Te$_5$/ TiW | >25 | – | 0.8 | 2 | 200 ns/ 30 ns | 0.57 pJ/ 14.17 pJ | [150] |
| MRL | VCM | TiN/HfO$_2$/Ti/TiN | 10 | 10$^5$ | <1.5 | <1.5 | 10 ns | 5.5 mW (read) / 11 mW (write) | [158] |

### 5.3.4. CRS used in IMPLY logic

Although CRS is primarily a non-stateful logic mechanism, Yang et al. demonstrated its use for IMPLY (stateful) logic.[131] The CRS array contains two types of CRS cells with different switching behaviors. The IMPLY operation (see also Figure 7, but the memristor is replaced by two cells like in CRS) is executed in three steps: 1) setting cell P to the ON state (both memristors in LRS), 2) performing the standard IMP operation via $V_{COND}$ and $V_{SET}$, and 3) resetting P (since it may be in state ON). This approach mitigates sneak path effects due to the CRS characteristics, but increases fabrication complexity. Each CRS cell requires a specialized stack, especially due to the different characteristics between individual CRS cells. Furthermore, the need for multiple programming steps adds latency and it is challenging to select the voltages for successful ON programming and IMP operations. Replacing PCM cells with CRS cells in a MAGIC-like architecture (Figure 8b) can theoretically enable CRS-based NAND, AND, NOR, and OR gates.[131] However, these designs remain confined to simulations, with no experimental verification reported to date.

### 5.3.5. V-R with PCM

Single-PCM device logic[132] also enables in-memory computation by carefully modulating the amplitude and duration of write pulses. V-R in this logic technique stands for voltage as input and resistance state as output.

This logic relies on the functionality of the PCM of additive crystallization, meaning that the cell can be crystallized in steps using the time dependency of the crystallization process and threshold switching of the amorphous state. An initial amorphous cell remains amorphous or is partially crystallized by two input pulses. Depending on their logical state, either low pulses (no crystallization) or higher pulses (partial crystallization) are applied. In a final "confirm" phase, the cell only then crystallizes fully, when a partial crystallization took place in the previous step.

Depending on the choice of crystallization time given by each input pulse, the logic functions NOR and NAND can be realized.

Despite its innovative simplicity, this method requires precise pulse control and repeated partial crystallization steps, highlighting the trade-offs between integration feasibility and operational complexity.

### 5.4. Non-Stateful - CMOS counterpart

#### 5.4.1. Memristor Ratioed Logic (MRL)

MRL employs a hybrid memristor-transistor logic technique based on a memristor configuration using two memristors in series with opposite polarities. By that, AND and OR logic gates (see **Figure 12a,b**) can be implemented. In MRL, both the inputs and the output are represented as voltages. When identical input voltages are applied, no voltage drop occurs across the memristors, leaving their resistance unchanged. Consequently, the output matches to the input voltage. For the OR gate, when the input voltages differ, the memristor corresponding to a logical "1" input switches from HRS to LRS, while the memristor associated with an input of logical "0" switches into HRS. The output is thus the voltage drop across the second memristor, resulting in a logical "1"; if either input is "1". Similarly, for the AND gate, due to the opposite polarity compared to the OR gate, the output is the voltage across the first memristor, resulting in a logical "0" if either input is "0". By integrating a CMOS inverter at the output, NOR and NAND gates can be implemented, achieving functional completeness and signal restoration.[133]

MRL enables the construction of standalone logic gates with a key advantage of increased logic density, as memristors are fabricated using metal layers stacked upon CMOS transistors. Furthermore, MRL seamlessly integrates with CMOS logic due to its voltage-based inputs and outputs. However, implementing MRL requires bipolar memristive elements, limiting suitable devices to RRAM and MRAM. Additionally, MRAM's low resistance ratio presents significant challenges for achieving reliable operation with MRL. Additionally, it is questionable whether MRL with MRAM devices increase logic density. Simulations using RRAM have demonstrated the feasibility of MRL gates, with recent studies analyzing failure mechanisms specific to MRL-based AND and OR blocks. Two primary failure types—cascading failures and race failures—have been identified and studied.[134] Additional modalities to make MRL crossbar-compatible have been proposed, such as CMOS pull-down networks integrated within crossbars[135] and full adders using transistors within crossbar arrays.[136] Recently, an experimental demonstration of MRL gate highlighted that voltage-dependent resistance states can limit functionality, emphasizing the need for further refinements.[137]

#### 5.4.2. 1T2M

The 1 Transistor 2 Memristor (1T2M) structure builds on the principles of MRL to implement additional logic gates, including the NOT gate (see Figure 12c).[138] In this design, the NMOS transistor in a conventional CMOS NOT gate is replaced by the MRL AND gate, allowing for the realization of various other gates such as XNOR. The 1T2M structure combines the advantages of MRL—such as leveraging memristors for compactness and integration above CMOS layers—with the versatility of CMOS logic, enabling a broader range of gate functionalities. This approach demonstrates potential for further enhancing the logic density of CMOS-based designs while incorporating emerging memory technologies for additional flexibility. Similar to MRL, cascading failures are a concern in the 1T2M structure.

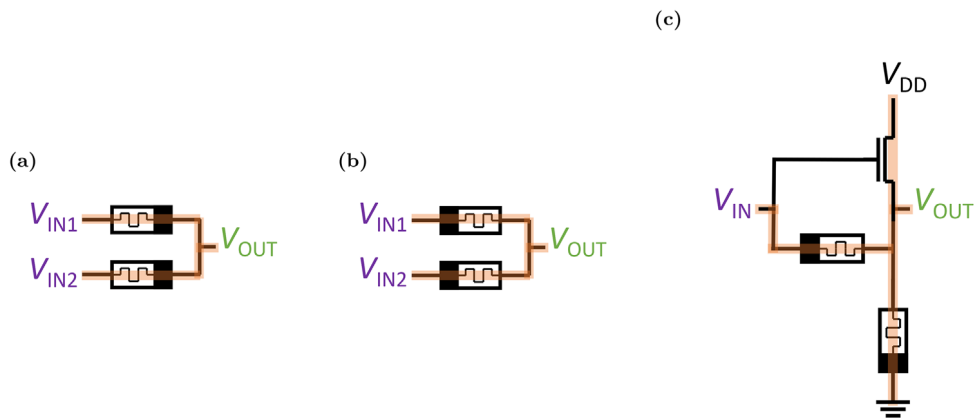

**Figure 12.** Non-stateful CMOS counterpart Memristor Ratioed Logic (MRL)[133]. Schematic of a) AND gate, b) OR gate, and c) 1T2M NOT gate.[138] The purple annotations show the logic input variables while the green annotations depict the logic output variable. The current path is highlighted in orange.

## 6. Comparison of Device Technologies and Logic Techniques

The previous sections outlined a variety of logic techniques, their operational principles, and the tradeoffs involved in implementing them using different emerging memory devices. While these individual analyses are informative, a direct comparison is essential to identify the relative strengths of each approach and determine their suitability for specific applications.

### 6.1. Framework for Comparison

To enable a meaningful comparison, it is important to define the evaluation criteria. A key distinction lies in classifying logic techniques as either *stateful* or *non-stateful*, a division that fundamentally impacts the required peripheral circuitry and overall system architecture. Additional parameters—such as parasitic effects, sense amplifier requirements, and array structures—further influence the performance, energy efficiency, and scalability of each logic technique, rendering the comparison inherently multidimensional.

Furthermore, the effectiveness of each logic technique is closely tied to the characteristics of the underlying device technology. Technologies such as RRAM, PCM, and MRAM exhibit distinct electrical and physical behaviors that shape logic functionality, reliability, energy efficiency, and integration potential. This section presents a comparative analysis that outlines the relative advantages and limitations of these techniques when paired with suitable memory devices.

### 6.2. Advantages and Disadvantages

**Table 2** summarizes the key features, strengths, and challenges of the reviewed logic techniques (detailed in Section 5) and their corresponding devices. Below is a qualitative breakdown:

### 6.3. Stateful Logic

#### 6.3.1. Advantages:

1. Enables computation directly within memory cells, significantly reducing data movement.
2. Highly compatible with dense memory arrays, offering strong scalability potential.

#### 6.3.2. Disadvantages:

1. Susceptible to input instability and logic drift.
2. Requires precise voltage control to maintain reliable operation.
3. Write-based computation can reduce device endurance due to frequent switching.

#### 6.3.3. Device Suitability:

1. RRAM and PCM are promising candidates, thanks to their high resistance ratios, though RRAM may suffer from device variability and endurance issues.
2. MRAM, despite its fast switching, high endurance, and low C2C variability, is challenged by low resistance contrast leading to a small read margin and less reliable logic operation, yet remains an active area of exploration.

### 6.4. Non-Stateful Logic

#### 6.4.1. Advantages:

1. Allows greater flexibility in circuit integration due to decoupled input and output representations.
2. Typically more tolerant to device variability, especially when logic is executed outside the memory array (e.g., near-memory implementations like Scouting Logic; see Section 5.2.1).

**Table 2.** Summary of the general advantages and disadvantages of the main logic techniques, comparing them to their device suitability. The device suitability is influenced by the results of experimental demonstrations and simulations, and each device characteristic. '+++' indicates most promising device integration without major known issues, '++' indicates functionality with minor drawbacks, '+' indicates general suitability but includes major challenges and '-' indicates low likelihood of possible integration.

| Logic Type | Logic | Device Suitability | | | | Advantages | Disadvantages |
|---|---|---|---|---|---|---|---|
| | | VCM | ECM | PCM | MRAM | | |
| Stateful | MAGIC | ++ | ++ | ++ | + | | |
| | | | | | | • Separated input & output | • Require high endurance |
| | IMPLY | ++ | ++ | + | + | | |
| | | | | | | • Energy efficient | • Input change<br>• High requirements on input voltages |
| Non-Stateful Near-Memory | Scouting | ++ | +++ | +++ | ++ | | |
| | | | | | | • No state change during logic operation → higher endurance | • SA necessary |
| Non-Stateful In-Memory | 1T1R | +++ | ++ | — | + | | |
| | | | | | | • Easy operation<br>• Only one cell used | • Specific array necessary |
| | CRS | +++ | ++ | — | — | | |
| | | | | | | • Overcoming sneak path issue | • Multiple cycles<br>• Destructive read out<br>• High voltages |
| Non-Stateful CMOS Counterpart | MRL | ++ | +++ | — | — | | |
| | | | | | | • Increasing logic density | • Not crossbar compatible |

### 6.4.2. Disadvantages:

1. Relies on additional peripheral circuitry, such as sense amplifiers and interconnects, which increases system complexity and can limit array density.
2. Often entails higher energy consumption due to greater data movement between memory and compute elements.

### 6.4.3. Device Suitability:

1. VCM excels in non-stateful logic due to its high resistance ratios and bipolar switching.
2. ECM devices offer precise switching behavior but suffer from limited endurance, constraining their broader applicability.
3. PCM is generally unsuitable (besides non-stateful near-memory) due to its unipolar switching and MRAM poses integration challenges due to large cell footprint—although novel circuit designs may help overcome these limitations.

## 6.5. Call for Action

While this comparison sheds light on the trade-offs and compatibilities between logic techniques and emerging memory devices, significant challenges remain in transitioning these concepts from laboratory demonstrations to scalable, real-world systems. Experimental validation is especially critical to assess the feasibility of many proposed combinations under practical conditions.

Although numerous promising simulation results exist in the literature, experimental demonstrations remain scarce. This gap can be attributed in part to the limited visibility and adoption of processing-in-memory approaches within the broader research community. Device development efforts often focus on standalone structures, and integrating these devices into crossbar arrays is both time-consuming and cost-intensive. Furthermore, newer device technologies still suffer from limited maturity, process variability, and insufficient device uniformity, which complicate the fabrication of reliable logic gates. Experimental implementation is further hindered by the high cost and complexity of customized measurement setups. Logic evaluation in crossbar

Table 3. Structuring technologies and logic techniques. Color coding indicates the maturity level of each approach: green denotes sufficient experimental demonstration; yellow indicates initial experimental results, but further investigation is needed; orange represents simulation-only evidence; and gray signifies that no work has been done to date.

| Device | Logic type | Logic | Input/Output Representation | R change of input | Demonstrated Gates | Initialization/ Read-out needed? | Year of Introduction | # of Max. Demonstrated Cycles |
|---|---|---|---|---|---|---|---|---|
| VCM | Stateful | MAGIC [94, 100] | $R_{M1}, R_{M2} / R_{M3}$ | no | NOR, AND, OR, NOT, NAND | yes / yes | 2014 [94] | 50 [97] |
| | | IMPLY [84, 94] | $R_{M1}, R_{M2} / R'_{M2}$ | yes | IMP, OR, NOR | yes / yes | 2010 [84] | $10^5$ [86] |
| | | stateful CRS [146] | $R_{CRSp}, R_{CRSq} / R'_{CRSq}$ | yes | IMP, NAND | yes / yes | 2014 [146] | – |
| | Near-Memory | Scouting [109, 112] | $R_{M1}, R_{M2} / I$ | no | AND, OR, XOR | yes / no | 2017 [109] | |
| | In-Memory | CRS [126] | $V_{T1,1}, V_{T1,2} / R_{CRS}$ | yes | 16 Boolean functions | yes / yes | 2010 [122] | $10^9$ [128] |
| | | 1T1R [119] | $V_G, V_{TE}, V_{BE}, R_{M1} / R'_{M1}$ | yes | 16 Boolean functions | yes / yes | 2017 [119] | $10^7$ [147] |
| | CMOS Counterpart | MRL [133] | $V_1, V_2 / V_{out}$ | – | AND, OR | no / no | 2012 [133] | 500 [137] |
| | | 1T2M [138] | $V_1, V_2 / V_{out}$ | – | NOT, XOR, XNOR | no / no | 2023 [138] | – |
| ECM | Stateful | direct resistance-coupling scheme [148] | $R_{M1}, R_{M2} / R_{M3}$ | no | 16 Boolean functions | yes / yes | 2022 [148] | 1 (one bit full-adder) [148] |
| | | IMPLY [84] | $R_{M1}, R_{M2} / R'_{M2}$ | yes | IMP, NOT, OR, COPY | yes / yes | 2010 [84] | 1 [90] |
| | Near-Memory | | | | | | | |
| | In-Memory | CRS [126] | $V_{T1,1}, V_{T1,2} / R_{CRS}$ | yes | 16 Boolean functions | yes / yes | 2010 [122] | 1 [126] |
| | CMOS Counterpart | | | | | | | |
| PCM | Stateful | PCM based stateful logic [105] | $R_{M1}, R_{M2} / R_{M3}$ or $R'_{M2}$ | yes | NOR, IMPLY, OR, NIMP | yes / yes | 2022 [105] | 50 [105] |
| | Near-Memory | Scouting PCM [114] | $R_{M1}, R_{M2} / I$ | no | AND, OR, XOR | yes / no | 2020 [114] | – |
| | In-Memory | Pinatubo [108] | $R_{M1}, R_{M2} / I$ | no | OR | yes / no | 2016 [108] | 100 [115] |
| | | V-R [132] | $V_{X1}, V_{X2} / R$ | – | NOR, NAND, NOT, XNOR | yes / yes | 2013 [132] | 1 [132] |
| | | [149] | $V_1, V_2 / R_M$ | – | AND, OR, NOT | yes / yes | 2013 [149] | 1 [149] |
| | | [150] | $V, B / R_M$ | – | AND, NAND, OR, NOR | yes / yes | 2016 [150] | 1 [150] |
| | CMOS Counterpart | | | | | | | |
| STT | Stateful | MAGIC [94], [100] | $R_{M1}, R_{M2} / R_{M3}$ | no | NOR | yes / yes | 2019 [100] | – |
| | | IMPLY [92] | $R_{M1}, R_{M2} / R_{M2}$ | yes | OR, AND, NIMP, IMP | yes / yes | 2018 [92] | 1 [92] |
| | | STT-CRAM [103] | $R_{M1}, R_{M2} / R_{M3}$ | no | NAND | yes / yes | 2020 [102] | – |
| | Near-Memory | HielM [117] | $R_{M1}, R_{M2} / I$ | – | – | – | 2018 [117] | – |
| | In-Memory | | | | | | | |
| | CMOS Counterpart | | | | | | | |
| SOT | Stateful | SpinPM [104] | $R_{M1}, R_{M2} / R_{M3}$ | no | NOR | yes / yes | 2019 [104] | |
| | | [105] | $R_{M1}, R_{M2} / R_{M3}$ | no | NOR, NAND, OR, AND | yes / yes | 2022 [105] | |
| | | V-SOTM [107] | – | – | AND, NAND, OR, NOR | – | 2021 [107] | |

(Continued)

 **e00348 (17 of 23)** 

**Table 3.** (Continued)

| Device | Logic type | Logic | Input/Output Representation | R change of input | Demonstrated Gates | Initialization/ Read-out needed? | Year of Introduction | # of Max. Demonstrated Cycles |
|---|---|---|---|---|---|---|---|---|
| | Near-Memory | PIMA-logic [118] | $R_{M1}$, $R_{M2}$ / V | no | AND, OR, NAND, NOR | yes / no | 2018 [118] | – |
| | In-Memory | [151] | – | – | AND, OR, NAND, NOR | yes / yes | 2003 [151] | – |
| | CMOS Counterpart | | | | | | | |

arrays often requires highly specific configurations and periphery, making scalability difficult. Moreover, discrepancies between real-world behavior and simulation arise due to incomplete modeling of device non-idealities, including variability, degradation, and parasitic effects—issues that are difficult to isolate experimentally.

Primary challenges in translating memristive logic techniques into scalable, commercial products include device reliability and quality, accurate device modeling, programmability, ease of manufacturing, and cost. Device reliability primarily addresses soft errors and the available hardware or software techniques to enhance robustness. Device quality consists of parameters such as resistance ratio, endurance, retention, and switching speed. For large-scale integration, precise device models that incorporate variations and aging effects are essential for reliable simulations. Additionally, the development of programming frameworks, as well as the ease of manufacturing and costs, remain critical challenges.

Critically, experimental validation is essential. The characterization of individual devices alone is insufficient for logic applications (recommended methods for single-device characterization are detailed in [80]). It is therefore necessary to demonstrate the endurance of logic operations, analyze reliability concerns, and verify the functionality of a statistically significant number of devices, with particular attention to reliability, like device-to-device (D2D) and cycle-to-cycle (C2C) variations. Investigating operational parameters, such as voltage ranges, to define stable operation windows would also be valuable. More detailed reliability analyses specific to logic functions would further advance the field.

For that, key research directions include:

1. Developing **standardized benchmarks** to facilitate cross-platform comparisons of logic techniques.
2. Exploring the **scalability of logic techniques** in large memory arrays under realistic workloads.
3. Investigating **hybrid approaches** that combine stateful and non-stateful logic for improved performance and adaptability.
4. Designing **innovative peripheral designs** to mitigate limitations posed by current sense amplifiers and array layouts.

This comparison framework serves as a guidepost for future efforts to co-optimize logic techniques and memory technologies,

ultimately advancing the development of efficient and reliable in-memory computing systems.

## 7. Conclusion

This review presents a comprehensive classification of logic families tailored for emerging memory technologies, detailing their switching mechanisms and the primary reliability challenges inherent to each device. By aligning logic families with compatible memory technologies, we summarized the state-of-the-art advancements and highlighted key reliability issues. **Table 3** and Table 1 compile prior work and provide a foundation for identifying promising directions for experimental validation using commercially viable materials.

Rather than asserting the superiority of any single logic technique, we acknowledge that optimal solutions are highly application-dependent. Our analysis emphasizes the relative strengths and limitations of different approaches, with particular attention to reliability concerns, statistical validation practices, and constraints in characterization and functionality. Experimental studies in RRAM devices, especially those employing valence change mechanisms (VCM), are currently the most prevalent. PCM devices have also demonstrated notable progress, while work on STT- and SOT-MRAM remains limited, reflecting ongoing challenges in their development.

The relevance of these memory technologies for processing-in-memory (PIM) applications is growing beyond academic research, with increasing industrial interest in their adoption. As Lanza et al.[49,139] observe, a significant disconnect persists between academia and industry. Academic efforts often focus on novel materials with exceptional, but impractical characteristics, while industry demands for scalable and manufacturable solutions. Many academic demonstrations emphasize high endurance or robust logic operation in single cells, which may not scale effectively in a commercial context. Main requirements for industrial adoption are fabrication costs, reliability, and ease of manufacturing particularly with respect to materials and fabrication tools while keeping new fabrication steps low. Additionally, advancing programming models and architectural support for memristor-based PIM operations will facilitate their integration into commercial products.[140] Since SRAM-based PIM techniques are already being explored in industry, non-stateful near-memory approaches, particularly Scouting Logic, are likely the easiest logic technique for commercial adoption.

Our findings reinforce that no single device or logic family excel across all evaluation criteria. Instead, selecting the appropriate technology requires careful consideration of application-specific priorities such as speed, energy efficiency, scalability, and reliability. The continued evolution of logic-in-memory systems will depend on bridging the academic-industrial divide, advancing device engineering, and validating scalable implementations.

Looking ahead, we anticipate progress in device reliability, logic integration density, and compatibility with commercial fabrication processes. This review aspires to serve as a roadmap for researchers, identifying critical gaps and encouraging deeper focus on experimental validation, practical deployment, and industry-aligned development of robust and versatile logic-in-memory systems.

## Acknowledgements

L.K. acknowledges support from the Israel Science Foundation (ISF grant 1397/24). P.K.A. acknowledges support from the U.S. National Science Foundation (award numbers 2400463, 2106562, and 2425538). S.K. acknowledges support from the European Research Council through the European Union's Horizon 2020 Research and Innovation Programme under Grant 757259. S.K. acknowledges support from the European Research Council through the European Union's Horizon Europe Research and Innovation Programme under Grant 101069336. S.K. acknowledges support from NSF-BSF grant number 2020-613.

## Conflict of Interest

The authors declare no conflict of interest.

## Author Contributions

T.N. and H.P. contributed equally to this work. All authors discussed and commented on the manuscript. The review paper was written under the leadership of S. Kvatinsky.

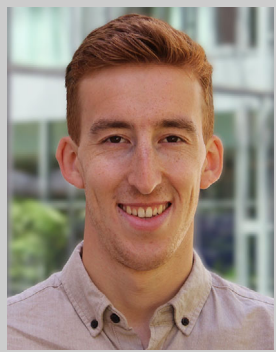

**Thomas Neuner** received his B.Sc. degree in electrical engineering from Technical University of Munich, Germany in 2022. He is currently pursuing his M.Sc. degree and is a visiting pre-doctoral fellow at Northwestern University, USA. His research interests include emerging memory technologies like resistive switching memories and spintronic-based memories and their applications toward in-memory computing and probabilistic computing.

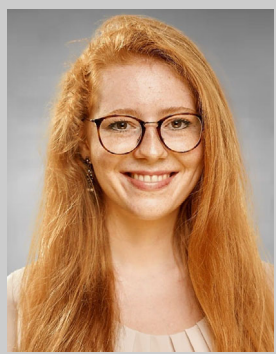

**Henriette Padberg** received the B.Sc. degree and M.Sc. degree in materials science from RWTH Aachen University, Germany in 2021 and 2024, respectively. Her current research focuses on the circuit design and experimental characterization of neuromorphic data converters using memristive devices.

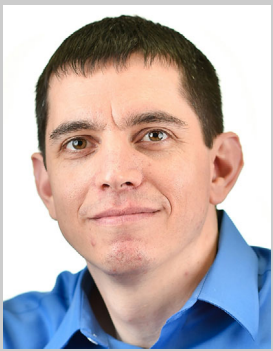

**Lior Kornblum** graduated with his PhD in Materials Science and Engineering at the Technion in 2012, followed by a postdoc at the group of Charles Ahn in Yale until 2015. Since 2016 he is a faculty member at the Viterbi Department of Electrical and Computer Engineering at the Technion, where he heads the Oxide Electronics Lab. Lior is a member of the Technion's Russell Berrie Nanotechnology Institute, the Grand Technion Energy Program, and the Resnick Catalysis Center.

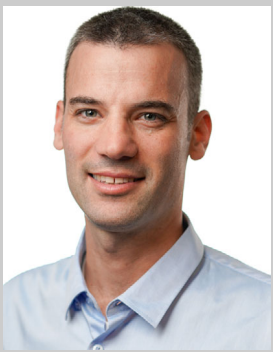

**Eilam Yalon** is an Associate Professor at the Andrew and Erna Viterbi Faculty of Electrical and Computer Engineering, Technion - Israel Institute of Technology. He received the B.Sc. degree in Materials Engineering and Physics in 2009 and the Ph.D. degree in Electrical Engineering in 2015, both from the Technion. From 2015 to 2018, he was a Post-Doctoral Researcher at Stanford University. His research focuses on improving the energy efficiency of nanoelectronics, with emphasis on memory and logic devices, 2D materials, ferroelectrics, and phase-change memory.

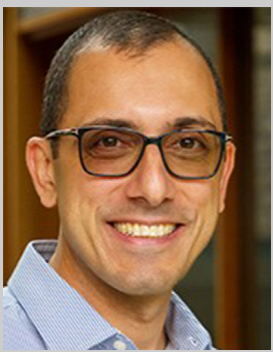

**Pedram Khalili Amiri** is a Professor of Electrical and Computer Engineering and Co-Director of the Applied Physics Program at Northwestern University. He received the PhD degree with distinction from Delft University of Technology (TU Delft). Prior to Northwestern, he was a postdoctoral researcher and adjunct assistant professor at the University of California, Los Angeles. He is a Distinguished Lecturer of the IEEE Nanotechnology Council and serves on the Executive Committee of the IEEE Task Force for Rebooting Computing. He is also a recipient of the Northwestern University ECE Allen Taflove Best Teacher Award. He is a Senior Member of the IEEE.

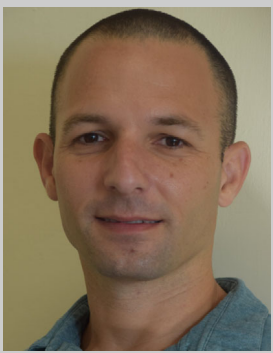

**Shahar Kvatinsky** is the Joan Goldberg Arbuse Chair Professor in Electronics at the Andrew and Erna Viterbi Faculty of Electrical and Computer Engineering, Technion – Israel Institute of Technology. Shahar received the B.Sc. degree in Computer Engineering and Applied Physics and an MBA degree in 2009 and 2010, respectively, both from the Hebrew University of Jerusalem, and the Ph.D. degree in Electrical Engineering from the Technion – Israel Institute of Technology in 2014. From 2006 to 2009, he worked as a circuit designer at Intel. From 2014 to 2015, he was a post-doctoral research fellow at Stanford University. Kvatinsky is a fellow of the Europe Young Academy, a former member of the Israel Young Academy, and the head of the Architecture and Circuits Research Center at the Technion.